# High-throughput Imaging and Spectroscopy of Individual Carbon Nanotubes in Devices with Light Microscopy

Kaihui Liu*[1], Xiaoping Hong*[1], Qin Zhou[1], Chenhao Jin[1], Jinghua Li[2], Weiwei Zhou[2], Jie Liu[2], Enge Wang[3], Alex Zettl[1,4], Feng Wang[1,4]

[1] Department of Physics, University of California at Berkeley, Berkeley, CA 94720, United States

[2] Department of Chemistry, Duke University, Durham, NC 27708, United States

[3] International Center for Quantum Materials, School of Physics, Peking University, Beijing 100871, China

[4] Materials Science Division, Lawrence Berkeley National Laboratory, Berkeley, CA 94720, United States

* These two authors contribute equally to this work

**Two paramount challenges in carbon nanotube research are achieving chirality-controlled synthesis and understanding chirality-dependent device physics[1-7]. High-throughput and in-situ chirality and electronic structural characterization of individual carbon nanotubes is crucial for addressing these challenges. Optical imaging and spectroscopy has unparalleled throughput and specificity[8-14], but its realization for single nanotubes on substrates or in devices has long been an outstanding challenge. Here we demonstrate video-rate imaging and in-situ spectroscopy of individual carbon nanotubes on various substrates and in functional devices using a novel high-contrast polarization-based optical microscopy. Our technique enables the complete chirality profiling of hundreds of as-grown carbon nanotubes. In addition, we in-situ monitor nanotube electronic structure in active field-effect devices, and observe that high-order nanotube optical resonances are dramatically broadened by electrostatic doping. This unexpected behaviour points to strong interband electron-electron scattering processes that can dominate ultrafast dynamics of excited states in carbon nanotubes.**

Single-walled carbon nanotubes (SWNTs) comprise a large family of tubular carbon structures characterized by different chiral indices (*n, m*), each having distinct electronic structure and physical properties[1]. They are promising materials for next generation nano-electronic and nano-photonic devices, including field-effect transistors, light emitters and photocurrent/photovoltaics device[1-7]. Currently nanotube research faces two outstanding challenges: (1) achieving chirality-controlled nanotube growth and (2) understanding chirality-dependent nanotube device physics. Addressing these challenges requires, respectively, high-throughput determination of nanotube chirality distribution on growth substrates and in-situ characterization of nanotube electronic structure in operating devices. Direct optical imaging and spectroscopy is well suited for these goals[8-14], but its realization for single nanotubes on substrates or in devices has been an outstanding challenge due to small nanotube signal and unavoidable environment background. Here we demonstrate for the first time high-throughput real-time optical imaging and broadband spectroscopy of individual nanotubes in devices using a polarization-based microscopy combined with supercontinuum laser illumination. Our technique is generally applicable to semiconducting and metallic nanotubes in various configurations, such as on (transparent or opaque) substrates, between contact electrodes, and under top gates. This is in contrast to strong constraints limiting other prevailing single-tube spectroscopy techniques: single-tube fluorescence spectroscopy only works for isolated semiconducting nanotubes[8]; Rayleigh scattering requires nanotubes suspended or oil-immersed on transparent substrate[9-12]; and resonant Raman scattering is limited by slow spatial and spectral laser scanning[15, 16].

Light polarization is extremely sensitive to minute optical anisotropy in a system, and has long been exploited to study materials ranging from molecules to crystals[17, 18].

Manipulation of polarization is especially suitable for carbon nanotube study because of the strong depolarization effect of one-dimensional nanotubes[19-21]. Figure 1 illustrates the scheme of our experimental design. From the interferometric point of view, optical contrast of a nanotube in a reflection configuration results from the interference between the nanotube-scattered electric field $E_{NT}$ and the substrate reflected electric field $E_r$ at the detector. This yields an optical contrast $\frac{\Delta I}{I} = \frac{|E_r + E_{NT}|^2 - |E_r|^2}{|E_r|^2} = \frac{2|E_{NT}|}{|E_r|}\cos\phi$, where $I$ is the optical signal from pure substrate reflection, $\Delta I$ is the optical signal difference resulted from the presence of a nanotube, and $\phi$ denotes the phase difference between $E_{NT}$ and $E_r$ at the detector. (The $|E_{NT}|^2$ term has been neglected because it is orders of magnitude smaller than the cross term.) Therefore one can, in principle, greatly enhance the nanotube contrast by reducing the reflection electrical field while maintaining the nanotube electrical field. This is achieved through polarization manipulation as shown in Fig. 1a. Horizontally polarized incident light (after polarizer $P_1$) illuminates a nanotube oriented at 45 degrees. The nanotube scattered electrical field ($E_{NT}^0$) is polarized along the nanotube direction (due to a strong depolarization effect on light perpendicularly polarized to nanotube), while the substrate reflection ($E_r^0$) retains the horizontal polarization, with $E_{NT}^0 / E_r^0 \sim 10^{-4}$ (Fig. 1b)[21]. The second polarizer ($P_2$) is oriented close to vertical direction (with a small angle deviation of $\delta$), which strongly reduces the reflection field to $E_r = E_r^0 \sin\delta$, but largely keeps the nanotube field $E_{NT} = E_{NT}^0/\sqrt{2}$. Therefore the nanotube contrast is enhanced by $1/(\sqrt{2}\sin\delta)$, which can reach over 100 with a reasonably small $\delta$ and produces $E_{NT}/E_r > \sim 10^{-2}$ (Fig. 1c).

Although the concept of polarization enhancement is appealingly simple, its experimental implementation to achieve wide-field imaging and spectroscopy of single nanotubes on substrates has been challenging due to constraints of optical microscopy and the polarization distortion caused by objectives[17]. Indeed, no individual carbon nanotubes on substrates can ever be observed in regular polarization microscopes. Recently a specialized polarization-based transmission microscopy was employed to probe single-nanotube optical absorption, but the technique is limited to suspended carbon nanotubes with no background from the substrate (as in previous Rayleigh scattering measurements[9-12]) and it requires slow frequency scanning of a Ti:sapphire laser[14]. Here we achieve wide-field imaging and high-throughput spectroscopy of individual nanotubes in devices by combining optimized polarization control and broadband supercontinuum illumination in reflection microscopy. The low coherence of broadband supercontinuum eliminates complications from laser interference and speckles, and at the same time it allows for broadband spectroscopy. In addition, we optimize the spatial mode of the high-brightness supercontinuum excitation to achieve the highest polarization purity together with wide-field illumination and high-resolution imaging. Consequently we obtain extinction ration orders of magnitude higher than that in regular polarization microscope, which is critical for single-tube imaging and spectroscopy on substrates.

Our technique enables direct imaging of single nanotubes in diverse configurations, as schematically illustrated in Fig. $2a_1$-$a_3$. Figure $2b_1$-$b_3$ show corresponding scanning electron micrographs, with a nanotube on a fused silica substrate ($2b_1$), a nanotube in a back-gated field-effect transistor with source-drain electrodes ($2b_2$), and a nanotube partly covered by an $Al_2O_3$ dielectric layer ($2b_3$). Now for the first time, we are able to image such individual nanotubes directly using optical microscopy (Fig. $2c_1$-$c_3$) rather than electron microscopy. A single-walled nanotube typically has a contrast larger than 5% in our optical microscopic images. The supplementary Movie S1 further demonstrates the capability of imaging single nanotubes on a substrate in real time. More importantly, we can not only "see" individual nanotubes, but also obtain their optical spectra and uniquely identify their chiralities. Figure $2d_1$-$d_3$ display the spectra of nanotubes shown in Fig. $2c_1$-$c_3$. Each spectrum is obtained within 2 seconds using the broadband supercontinuum illumination and a spectrometer equipped with a linear array charge coupled device (CCD). From the prominent optical resonances in the spectra, we can assign the chirality (20, 6), (22, 16), and (26, 22) to these three SWNTs[10]. They are semiconducting, metallic and semiconducting nanotubes with diameters of 1.8, 2.6 and 3.3 nm, respectively.

Such high-throughput imaging and chirality identification of nanotubes on substrates can be an indispensable tool for improving carbon nanotube growth. The Holy Grail in nanotube synthesis, the chirality control, requires systematic optimization of nanotube growth conditions. A critical component in the growth optimization is the feedback from characterization on nanotube chirality distribution from different growth conditions. This component has been missing so far because there is no simple and reliable way to accurately determine detailed nanotube species and abundance (except for a rough diameter distribution using resonant Raman measurement). Our technique here can readily image hundreds of as-grown nanotubes on substrates and further determine their chirality with high throughput. In Fig. 3 we plot the chirality of over 400 SWNTs from one growth condition (See Methods for growth details), which includes 240 semiconducting nanotubes (S) and 162 metallic nanotubes (M). For the first time detailed chirality distribution of hundreds of nanotubes on as-grown substrates (transparent or opaque) can be accurately determined. The chirality distribution in Fig. 3a shows that in the specific sample semiconducting and metallic nanotubes are enriched in different region with characteristic chiral angle and diameter dependence. The chiral angle distribution (Fig. 3c) shows that large chiral angles (close to the armchair direction) are more favorable, consistent with previous findings[8]. The diameter distribution (Fig. 3b), however, is quite surprising. It reveals a strong correlation between the diameter and a nanotube being semiconducting or metallic. With random distribution of chirality, one expects a ratio of 2:1 for semiconducting and metallic nanotubes. However, we observe that for nanotube diameter between 1.7-2.1 nm, semiconducting species are highly enriched, while metallic ones completely dominate for nanotube diameter larger than 2.3 nm. This unusual correlation behaviour can only be revealed with our capability to map all individual nanotube chirality on substrates. In comparison, previous Raman characterization and electrical measurements only shows an overall enrichment of semiconducting nanotubes in this growth condition without observing any diameter-metallicity correlation[22], because Raman scattering selectively probes only nanotube species in resonance with the excitation laser and can miss the full picture. The accurate and complete characterization of as-grown nanotube species enabled by our technique will be crucial for better understanding of the growth mechanisms and systematic growth optimization.

In-situ imaging and spectroscopy of individual nanotubes also offer new opportunities to probe nanotube physics in operating devices. Here we examine gate-variable nanotube optical transitions in field-effect devices (Fig. 2a$_2$) to investigate electron-electron interaction effects on excited states in nanotubes.

Figure 4a and 4b display gate-dependent optical spectra for an (18, 18) metallic nanotube and a (26, 10) semiconducting nanotube, respectively. The resonance peak in metallic (18, 18) nanotube arise from $M_{22}$ transitions, and the peaks in semiconducting (26, 10) are from $S_{44}$ and $S_{55}$ transitions. All these optical resonances show significant broadening with gate voltage varying from close to 0 V to -30 V, which corresponds to a nanotube doping from the charge neutral point to a hole density of ~ 0.45 e/nm (based on calibration using G-mode Raman resonance in metallic nanotubes[23]; See supplementary information S1). At such doping levels, free holes partially fill the linear band of metallic nanotubes or the first subband of semiconducting nanotubes. Therefore the broadening of higher-band optical transition cannot be accounted by Pauli blocking that dominates semiconducting nanotube fluorescence or graphene absorption[24, 25]. Instead, it originates from many-body interactions between doped carriers and excitons in carbon nanotubes.

Electron-electron interactions can be greatly enhanced in one dimension. Two types of interactions between doped carriers and excitons were well known to affect excitonic resonances in nanotubes: dielectric screening[26, 27] and formation of trion states[28]. Trions will lead to a new optical resonance[28], which we do not observe in the higher subband transitions. Dielectric screening of the nanotube exciton is expected to shift the exciton transitions with no increase in the optical width[26], and previous Raman studies suggest that this dielectric screening dominates gate-induced effects on excitonic transitions[27]. In our experiment, we do observe a small redshift in most exciton transitions, presumably due to the screening effect. However, the gate-induced effect is dominated by a broadening of exciton resonance width. It indicates that a new type of electron-exciton interaction is critically important.

Because inhomogeneous broadening in single-tube spectra is small, the resonance width can be directly related to the ultrafast dynamics of an excited state. The observed gate-induced broadening cannot be from a change in dephasing due to exciton-phonon coupling, because electrical gating mainly introduces free carrier doping with little effect on exciton-phonon interactions.

Here we propose that inter-subband scattering between the exciton and gate-induced free carriers (Fig. 4c, 4d and supplementary information S2) could be responsible for the ultrafast dephasing of exciton through population decay. This electron-electron scattering is an Auger-type process, and it is strongly constrained in one-dimensional carbon nanotube by the conservation of energy, momentum, and angular momentum (described by the quantum number $E$, $k$, and band index $\mu$, respectively[1]). Fig. 4c (4d) shows one representative scattering channel that satisfies the stringent conservation requirements in hole doped metallic (semiconducting) nanotubes. Such scattering between optically excited electron and free holes (in another valley) is absent in pristine undoped carbon nanotube (Fig. $4c_1$ and $4d_1$), and emerges with hole doping (Fig. $4c_2$ and $4d_2$). (Excitonic correlation between the excited electron and hole, not shown in the illustration for simplicity, should not change the picture qualitatively.) This inter-subband electron-electron scattering rate increases with the free carrier concentration, and can dominate ultrafast relaxation of the exciton state at high doping.

It is interesting to note that carbon nanotube provides a unique opportunity to probe different ultrafast processes in graphitic materials, which share similar electronic structures and dynamic responses. Unlike graphene, the well-defined exciton resonances in nanotubes allow one to estimate the ultrafast excited state dynamics from resonance widths using Heisenberg uncertainty principle. In addition, the stringent constraint from energy, momentum, and angular momentum conservation means that excitons can relax only through electron-phonon interactions in undoped nanotubes, and allows us to isolate its contribution to ultrafast dynamics. On the other hand, gate dependence probes selectively the exciton relaxation through electron-electron interactions. Take the $M_{22}$ transition in (18, 18) nanotube for an example. We can isolate a dephasing rate of $\sim (5 \text{ fs})^{-1}$ from exciton-phonon coupling and a decay rate of $\sim (10 \text{ fs})^{-1}$ from exciton-electron coupling at doping level of ~0.45 e/nm. Such knowledge on ultrafast relaxation will be important for optoelectronic applications employing hot electrons in carbon nanotube and graphene[29, 30].

In summary, our polarization-based optical microscopy enables for the first time a high-throughput imaging and spectroscopy of individual carbon nanotubes on substrates and in devices. In addition to carbon nanotube, the technique can also greatly enhance optical contrast of other "invisible" anisotropic materials. This capability can open up exciting opportunities in studying a variety of one-dimensional nanomaterials, such as graphene nanoribbons, semiconductor nanowires and nanorods, and nano-biomaterials like actin filaments.

**Methods:**

**Growth of carbon nanotubes on different substrates**

Long nanotubes are grown by chemical vapor deposition (CVD) on various substrates, including fused silica, quartz, $SiO_2$/Si, $Si_3N_4$/Si and $Al_2O_3$/$SiO_2$/Si substrates. For the study of nanotube device, we typically use methane in hydrogen ($CH_4$:$H_2$=1:2) as gas feedstock and a thin film (~ 0.2 nm) of iron as the catalyst on $SiO_2$/Si substrate for CVD growth at 900 ºC[31]. This growth condition yields nanotubes with spacing of tens of microns. For the study of chirality distribution, we grow the nanotubes on Y-cut single crystal quartz substrates at 900 ºC[22]. In this growth, we used ethanol plus water through Ar bubble (90 sccm for ethanol and 30 sccm for water at 0 ºC) in hydrogen (280 sccm) as gas feedstock, and used $CuCl_2$/poly(vinylpyrrolidone) alcohol solution as catalysts.

**Fabrication of carbon nanotubes devices**

Back-gated nanotube field-effect transistor in Fig. $2a_2$ was fabricated on 90 nm $SiO_2$/Si substrate. The two gold electrodes (20 nm thick) were evaporated on the nanotube by electron beam evaporation. The $Al_2O_3$ layer (15 nm thick) in Fig. $2a_3$ was evaporated on nanotube by electron beam evaporation.

**Optical measurements**

A supercontinuum laser (470 nm ~ 1800 nm) is used as the light source[9, 10]. A reflective microscope is used, where the objective serves to focus the supercontinuum light to the sample and then collect the nanotube scattering and substrate reflection light. One polarizer is placed in the incident beam with its transmission axis set horizontally. A second polarizer is placed in the detection beam, and its polarization is controlled to be at a small angle $\delta$ to the vertical direction. For nanotube imaging the incident beam is focused to about 30 microns in diameter, which sets the field of view. The image is taken by a consumer Nikon D5100 camera with an integration time of ~ 20 ms. The video is taken at a rate of 30 frames per second. For nanotube spectroscopy we focus the supercontinuum down to ~1 micron by expanding the incident beam, and analyze the outgoing radiation with a spectrograph equipped with an array CCD detector. Two sets of spectra with the nanotube inside beam ($I_{inside}$) and outside beam ($I_{outside}$) are taken and the final spectrum is obtained as $\Delta I/I=(I_{inside}-I_{outside})/I_{outside}$.

**Nanotube chirality assignment from optical resonances**

For each single-walled carbon nanotube spectrum, transition energies at the optical resonance peaks are identified between 1.4 and 2.6 eV. We compare these transition energies to the atlas of suspended nanotube optical transitions to assign the nanotube chirality[10]. To account for the dielectric screening effect from the substrate, a 40 meV redshift is added to the transition energies from the atlas[32]. For about 90% single-wall nanotubes with diameter between 0.8-3.4 nm, the identification is unique. The others have two or three possible assignment. However, all these possible candidates typically belong to the same family with nearly identical diameter. Their physical properties are therefore very similar and this assignment uncertainty will not affect most nanotube studies.

**Acknowledgements**: This study was supported by NSF CAREER grant (No. 0846648), the NSF Center for Integrated Nanomechanical Systems (No. EEC-0832819), and NSF CHE-1213469. Support for characterization instrumentation was provided by the Director, Office of Energy Research, Materials Sciences and Engineering Division, of the US Department of Energy under Contract No. DE-AC02- 05CH11231. J.L. and W.Z. also acknowledge the support from Duke SMiF (Shared Materials Instrumentation Facilities).

**Figure captions**

**Figure 1. Scheme of polarization-based optical microscopy for single-nanotube imaging and spectroscopy.** (**a**) Configuration of the incident polarizer $P_1$, outgoing polarizer $P_2$, and the carbon nanotube. (**b-c**) Illustration of electrical field polarization before (b) and after (c) polarizer $P_2$. Dramatic reduction of the reflection electrical field leads to an enhancement of nanotube optical contrast by $\sim 1/(\sqrt{2}\sin\delta)$, where $\delta$ is the deviation angle of $P_2$ from perpendicular direction.

**Figure 2. Optical imaging and spectroscopy of individual nanotube on substrates and in devices.** (**$a_1$-$a_3$**) Schematic drawing of a nanotube on fused silica substrate, a nanotube in a back-gated field-effect transistor device (with two gold electrodes), and a nanotube partly under $Al_2O_3$ dielectric layer. (**$b_1$-$b_3$**) Scanning electron microscopic (SEM) images of the nanotubes corresponding to $a_1$-$a_3$. The dashed line in $b_3$ traces the invisible $Al_2O_3$ edge. Scale bars are 5 microns. (**$c_1$-$c_3$**) Direct optical images of individual nanotubes in $b_1$-$b_3$ using a color camera with an integration time of $\sim$ 20 ms. Optical images show high contrast for all individual nanotubes. (**$d_1$-$d_3$**) Optical spectra of the nanotubes in $c_1$-$c_3$, from which we identify the three nanotubes with chirality of (20, 6), (22, 16), and (26, 22). They are semiconducting, metallic and semiconducting nanotubes with diameters of 1.8, 2.6 and 3.3 nm, respectively.

**Figure 3. High-throughput determination of 402 single-walled carbon nanotube chirality distribution from one growth condition.** (**a**) Chiral index distribution of semiconducting (red triangles) and metallic nanotubes (dark green circles) which show

enrichment in different n-m region. **(b)** Diameter distribution of semiconducting (red bars) and metallic nanotubes (dark yellow bars). They reveal a surprising correlation: semiconducting species are highly enriched for nanotube diameter between 1.7-2.1 nm, while metallic species dominate for tube diameter larger than 2.3 nm. **(c)** Chiral angle distribution of semiconducting and metallic nanotubes. Both show that large chiral angles (close to the armchair direction) are more favorable, consistent with previous findings[8].

**Figure 4**. **Gate-variable nanotube optical transitions in field-effect devices.** (**a**)(**b**) Optical spectral evolution of a metallic (18, 18) (a) and a semiconducting (26, 10) nanotube (b) under different back-gated voltages. A significant broadening was observed for all transition from $V_g = \sim 0$ V to $V_g = -30$ V, which correspond to changes from undoped state to hole doping of $\sim 0.45$ e/nm. At this doping level, free holes partially fill the linear band of metallic nanotubes or the first subband of semiconducting nanotubes. (**c**)(**d**) Schematic illustration of one representative ultrafast decay pathway of the optically excited electron due to inter-subband electron-electron scattering in doped metallic (c) and semiconducting (d) nanotubes. (See supplementary information for the other three related decay pathways). The photo-excited electron decays to a state in a lower subband and transfers its energy, momentum and angular momentum to a free hole in the other valley. This process requires a free hole to participate, and it is forbidden for undoped nanotubes ($c_1$ and $d_1$) but allowed for doped ones ($c_2$ and $d_2$). The insets in c and d depict the process in two-dimensional graphene Brillouin zone, where the vertical lines correspond to nanotube subbands (their distance is exaggerated for better vision). It shows that energy, momentum and angular momentum are conserved in this inter-subband electron-electron scattering.

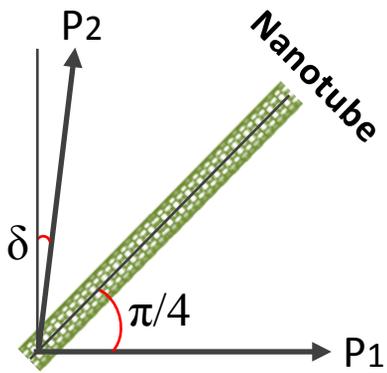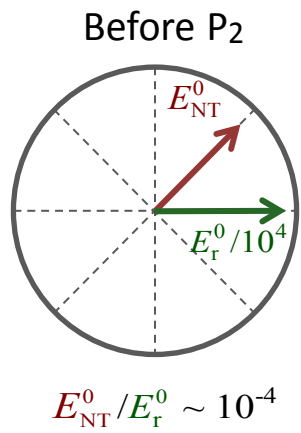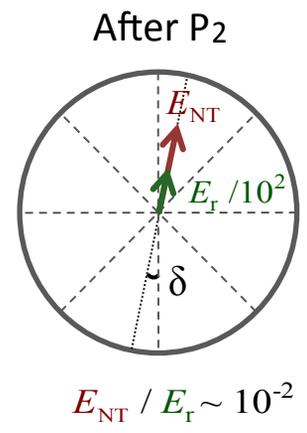

**Figure 1**

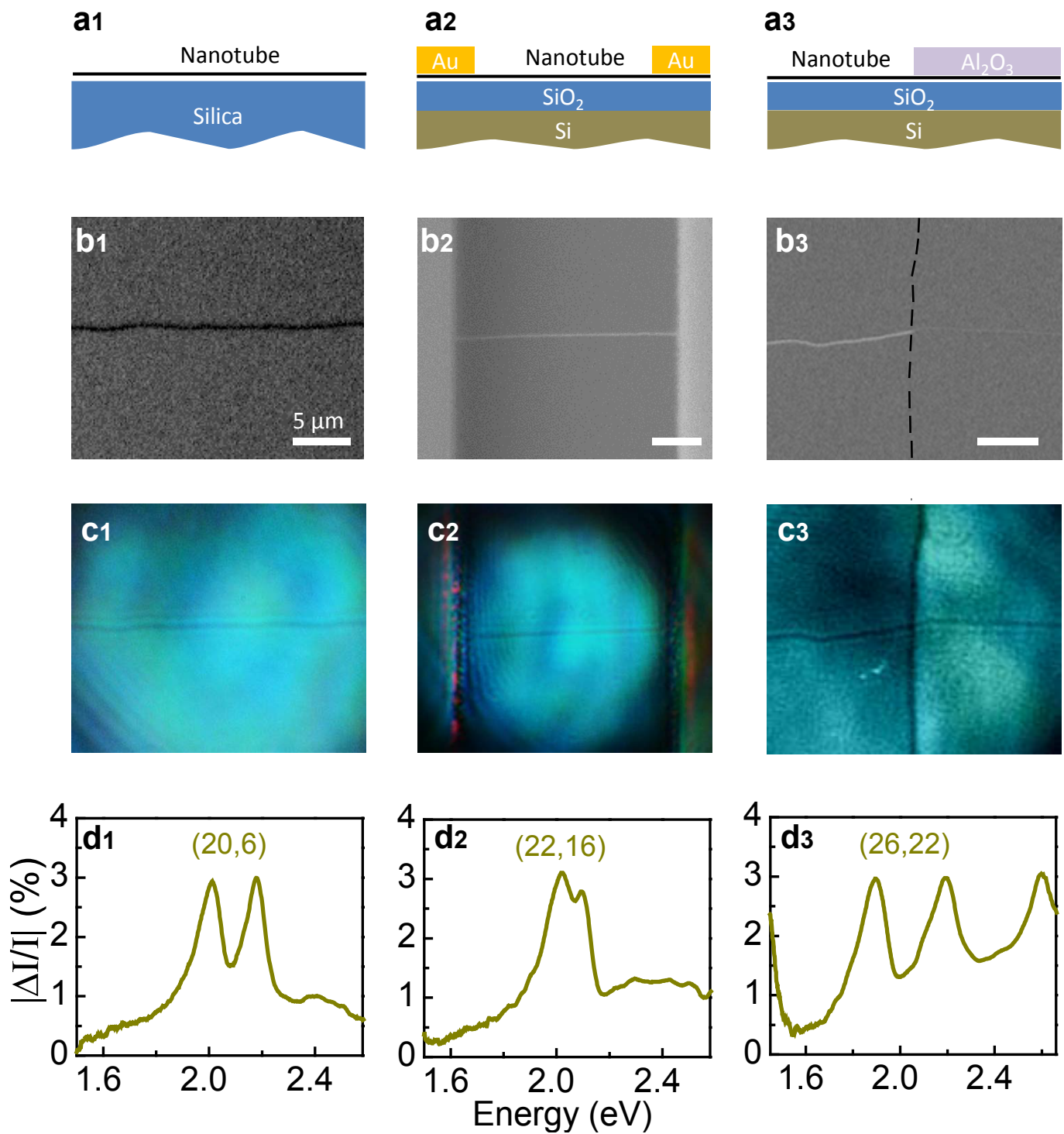

**Figure 2**

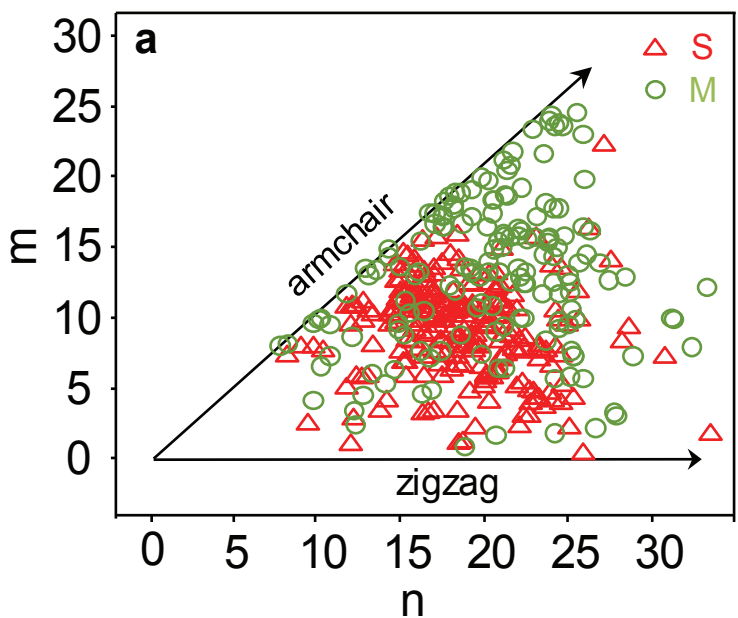
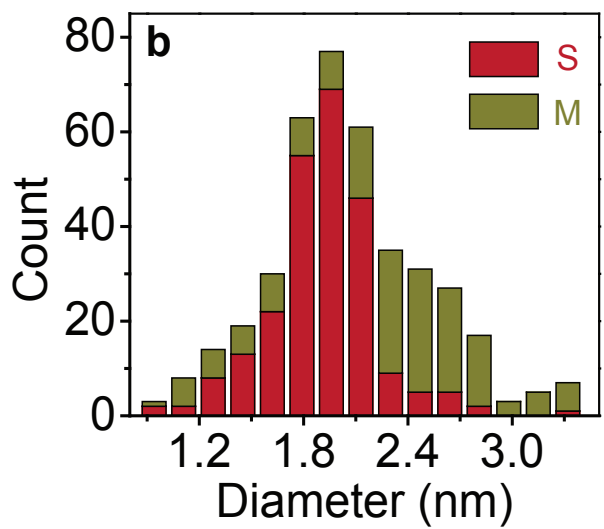
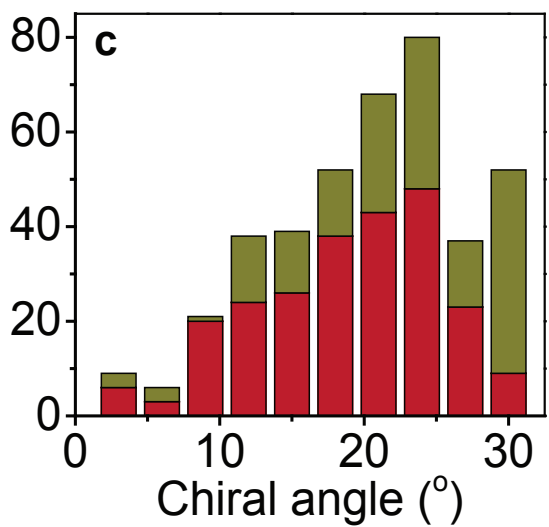

**Figure 3**

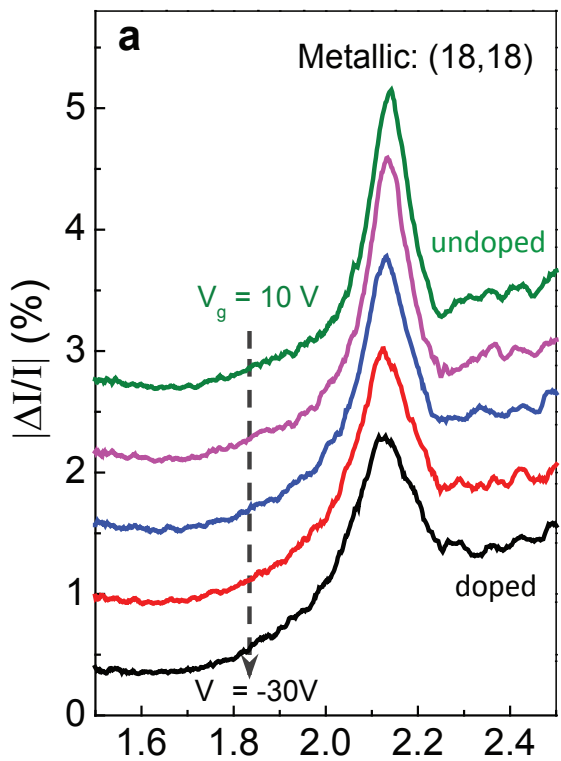
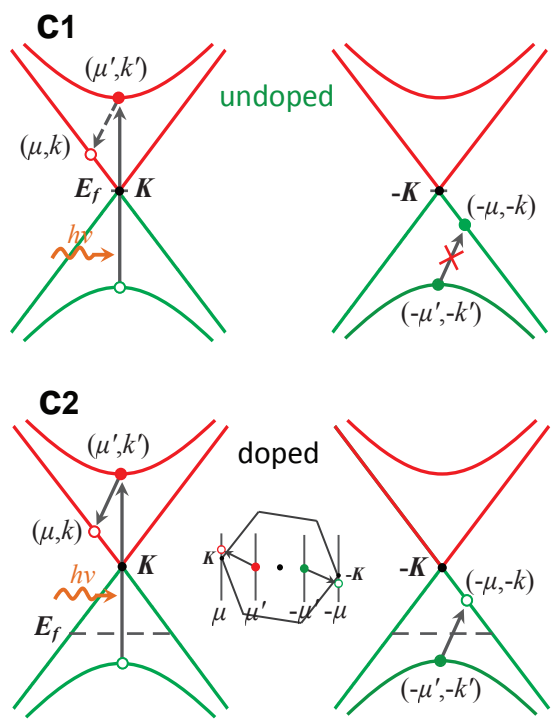
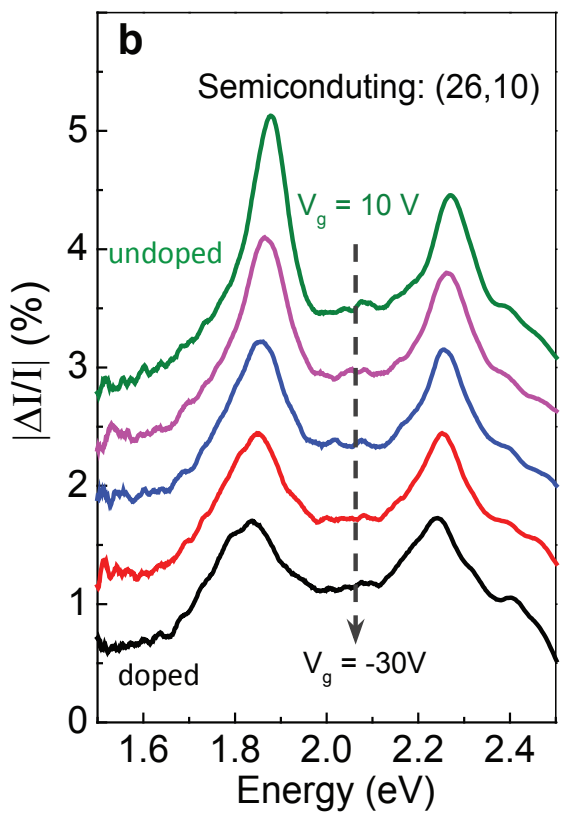
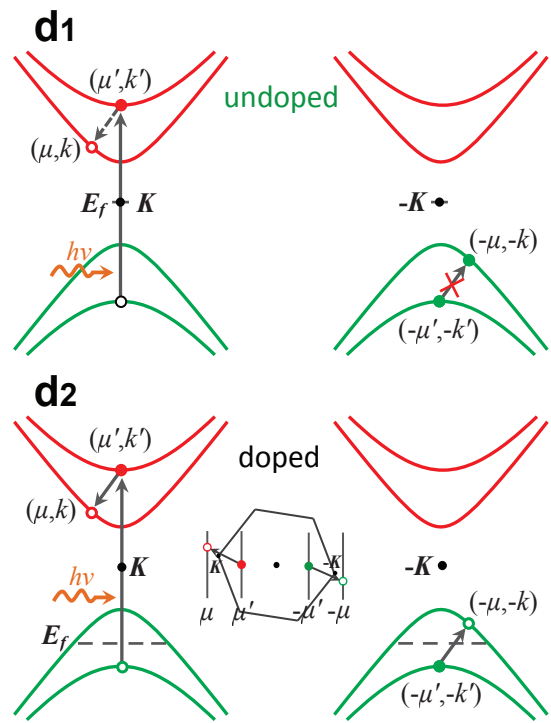

**Figure 4**

Supplementary information for

# High-throughput Imaging and Spectroscopy of Individual Carbon Nanotubes in Devices with Light Microscopy

Kaihui Liu*[1], Xiaoping Hong*[1], Qin Zhou[1], Chenhao Jin[1], Jinghua Li[2], Weiwei Zhou[2], Jie Liu[2], Enge Wang[3], Alex Zettl[1,4], Feng Wang[1,4]

[1] Department of Physics, University of California at Berkeley, Berkeley, CA 94720, United States

[2] Department of Chemistry, Duke University, Durham, NC 27708, United States

[3] International Center for Quantum Materials, School of Physics, Peking University, Beijing 100871, China

[4] Materials Science Division, Lawrence Berkeley National Laboratory, Berkeley, CA 94720, United States

**S1. Gating efficiency calibration using G-mode Raman resonance in metallic nanotubes**

**S2. Inter-subband free hole-excited electron scattering channels in doped nanotubes**

**S3. Supplementary Movie caption**

**S1: Gating efficiency calibration using G-mode Raman resonance in metallic nanotubes**

We calibrated the gating efficiency of our field-effect device by characterizing the gate-dependent G-mode Raman peak in non-armchair metallic carbon nanotubes[1]. Figure S1 shows one calibration example from a (13, 1) nanotube on 90nm $SiO_2$/Si substrate with back-gate field-effect transistor geometry. Doping in metallic nanotube blocks the low energy electronic transitions, and it changes the coupling between electrons and the longitudinal optical (LO) phonon. A quantitative fitting of this effect on G-mode Raman peak shift and linewidth change for (13, 1) nanotube are presented in Fig. S1b and S1c, respectively. From the theoretical fitting, we can deduce a gating efficiency of 0.013 hole/(nm.V) for line charge density in this (13, 1) nanotube (diameter =1.1 nm).

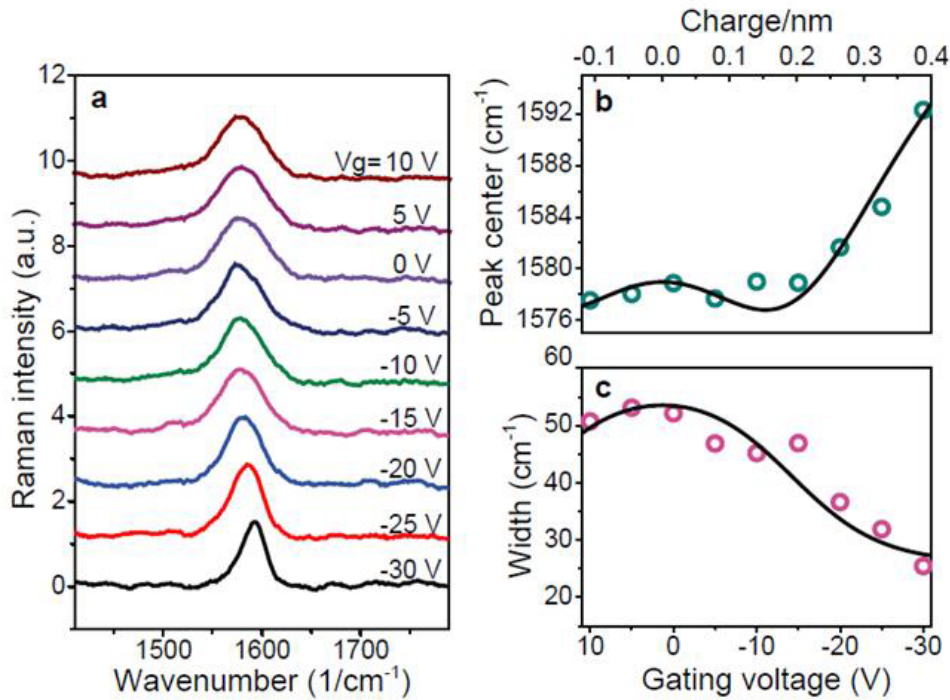

Figure S1. (**a**) G-mode Raman spectra for (13, 1) metallic nanotube at different back-gated voltages. (**b**)(**c**) Changes in the Raman peak center (b) and width (c) as a function of gating voltages (dots). Solid lines are based on fitting using the model in Ref. 1, which relate the G-mode Raman to carrier concentration in the metallic nanotube (top axis). Comparison between the experiment and theory yields the gating efficiency of this nanotube field-effect transistor.

For nanotubes of different diameter, the classic capacitance scales with $1/\ln(4t/d)$, where $t$ is thickness of dielectric layer and $d$ is the nanotube diameter. Therefore this gating efficiency varies with the diameter only weakly in a logarithmic dependence. (At the relatively high gating voltages used here, quantum capacitance of the nanotube is not important[2]). We calibrated three metallic nanotubes and found that the gating efficiencies are similar in these devices, with a variation less than 30% after accounting for the diameter dependent capacitance.

**S2: Schematics for all four different inter-subband scattering channels between free holes and optical excited electrons in a doped metallic nanotube that can contribute to the ultrafast dephasing of the exciton.**

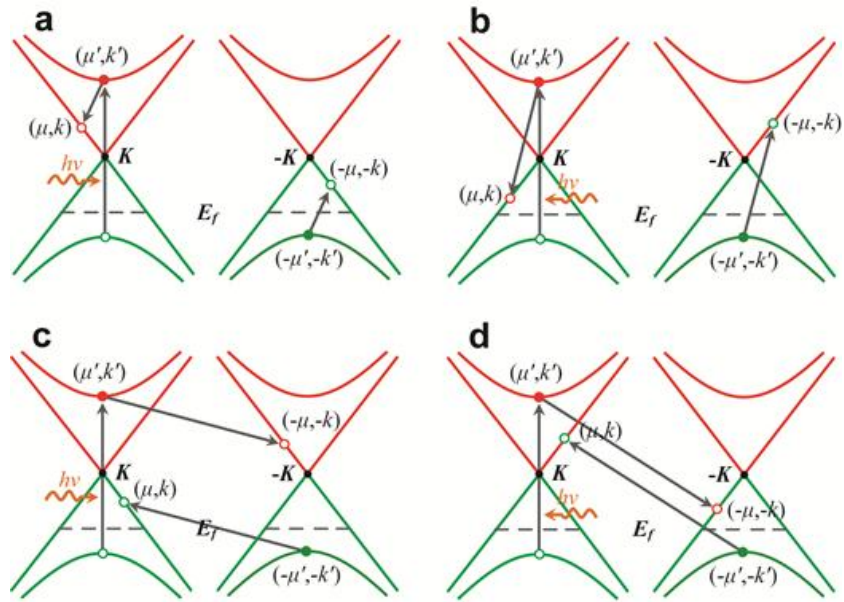

Figure S2. Four representative electron scattering channels that satisfy the energy, momentum and angular momentum conservation requirement in doped metallic nanotubes

The picture is quite similar for semiconducting nanotubes, except that the lowest band is parabolic instead of linear as in metallic nanotube.

## S3. Supplementary Movie caption

Real-time video of tracing a carbon nanotube sample on $SiO_2$/Si substrate using our polarization-based homodyne microscopy. The view size is ~30 microns in diameter.

- 00:01: Upper-right part of the nanotubes is under $Al_2O_3$ layer; Lower-left part of the nanotubes is on $SiO_2$/Si substrate.
- 00:39: A few different nanotubes appear.
- 02:10: Another nanotube perpendicular to the first one is seen with opposite contrast, as predicted by the polarization-based homodyne method when the nanotube orientation changes from 45 to 135 degrees.
- 02:27 Upper-right part of the nanotubes is on $SiO_2$/Si substrate; Lower-left part of the nanotubes is under $Al_2O_3$ layer.